\documentclass[12pt]{article}

\usepackage{amssymb}
\usepackage{amsmath}
\usepackage{amsbsy} % bold Greek etc.
\usepackage{graphics}

\begin{document}

\title{\bf RS model with a small curvature and gravity effects in $\mathbf{e^+e^-}$ annihilation
into leptons at the LC}

\author{A.V. Kisselev\thanks{Electronic address:
alexandre.kisselev@ihep.ru} \\
\small Institute for High Energy Physics, 142281 Protvino, Russia}

\date{}

\maketitle

\thispagestyle{empty}

\bigskip

\begin{abstract}
The Randall-Sundrum (RS) model with a small curvature is considered.
The mass spectrum of Kaluza-Klein (KK) gravitons in such a scheme is
similar (although not equivalent) to that in a model with one extra
dimension in a flat metric. The gravity effects in the processes
$e^+e^- \rightarrow e^+e^-$ and $e^+e^- \rightarrow\mu^+\mu^-$ at
the collision energy 1 TeV are presented. Our calculations are based
on the previously obtained formula for virtual graviton
contributions which takes into account both a discrete character of
the mass spectrum and nonzero widths of the KK gravitons.
\end{abstract}

%%%%%%%%%%%%%
% Main text %
%%%%%%%%%%%%%

\section{Randall-Sundrum model with the small curvature}
\label{sec:RS_model}

The Randall-Sundrum (RS) model~\cite{Randall:99} is one realization
of extra dimension (ED) theories in a slice of the AdS$_5$
space-time with the following background warped metric:
\begin{equation}\label{metric}
ds^2 = e^{2 \kappa (\pi r - |y|)} \, \eta_{\mu \nu} \, dx^{\mu} \,
dx^{\nu} + dy^2 \;,
\end{equation}
where $y = r \theta$ ($-\pi \leqslant \theta \leqslant \pi$), $r$
being the ``radius'' of extra dimension, and  $\eta_{\mu \nu}$ is
the Minkowski metric. The points $(x_{\mu},y)$ and $(x_{\mu},-y)$
are identified, so we work on the orbifold $S^1/\mathbf{Z_2}$.

The parameter $\kappa$ defines a 5-dimensional scalar curvature of
the AdS$_5$ space. Namely, the Ricci curvature invariant for this
AdS$_5$ space, $\mathcal{R}^{(5)}$, is given by $\mathcal{R}^{(5)} =
-20 \, \kappa^2$. For the sake of simplicity, in what follows, we
will call $\kappa$ ``curvature''.

The model has two 3D branes with equal and opposite tensions located
at the point $y = \pi r$ (called the \emph{TeV brane}, or
\emph{visible brane}) and point $y = 0$ (referred to as the
\emph{Plank brane}, or \emph{invisible brane}). If $k > 0$, then the
tension on the TeV brane is negative, whereas the tension on the
Planck brane is positive. All the SM fields are constrained to the
TeV brane, while the gravity propagates in five dimensions.

It is necessary to note that the metric \eqref{metric} is chosen in
such a way that 4-dimensional coordinates $x_{\mu}$ are Galilean on
the TeV brane where all the SM field live, since the warp factor is
equal to unity at $y = \pi r$.%
\footnote{To get a right interpretation, one has to calculate the
masses on each brane in the Galilean coordinates with the metric
$g_{\mu \nu} = (-1,1,1,1)$~\cite{Rubakov:01,Boos:02} (see below for
details).}

By integrating a 5-dimensional action over variable $y$, one gets an
effective 4-dimensional action, that results in the ``hierarchy
relation'' between the reduced Planck scale $\bar{M}_{\mathrm{Pl}}$
and 5-dimensional gravity scale $\bar{M}_5$:
\begin{equation}\label{RS_hierarchy_relation}
\bar{M}_{\mathrm{Pl}}^2 = \frac{\bar{M}_5^3}{\kappa} \left( e^{2 \pi
\kappa r} - 1 \right) \;.
\end{equation}
The reduced 5-dimensional Planck mass $\bar{M}_5$ is related to
the Planck mass by $M_5 = (2\pi)^{1/3} \bar{M}_5 \simeq 1.84
\bar{M}_5$ (see, for instance, \cite{Giudice:99}).

From the point of view of a 4-dimensional observer located on the
TeV brane, there exists an infinite number of graviton KK
excitations with masses
\begin{equation}\label{graviton_masses}
m_n = x_n \, \kappa, \qquad n=1,2 \ldots \;,
\end{equation}
where $x_n$ are zeros of the Bessel function $J_1(x)$.

The interaction Lagrangian on the visible brane looks like the
following:
\begin{equation}\label{Lagrangian}
\mathcal{L} = - \frac{1}{\bar{M}_{\mathrm{Pl}}} \, T^{\mu \nu} \,
G^{(0)}_{\mu \nu} - \frac{1}{\Lambda_{\pi}} \, T^{\mu \nu} \,
\sum_{n=1}^{\infty} G^{(n)}_{\mu \nu} +
\frac{1}{\sqrt{3}\Lambda_{\pi}} \, T^{\mu}_ {\mu} \, \phi \;.
\end{equation}
Here $T^{\mu \nu}$ is the energy-momentum tensor of the matter on
this brane, $G^{(n)}_{\mu \nu}$ is a graviton field with the
KK-number $n$, and $\phi$ is a scalar field called radion. The
parameter $\Lambda_{\pi}$ in Eq.~\eqref{Lagrangian},
\begin{equation}\label{lambda}
\Lambda_{\pi} = \bar{M}_5 \,\left( \frac{\bar{M}_5}{\kappa}
\right)^{\! 1/2} ,
\end{equation}
is a physical scale on the TeV brane.

In most of the papers which treat the RS model,%
\footnote{Including the original one \cite{Randall:99}.}
the background metric is taken to be
\begin{equation}\label{metric_uncorr}
ds^2 = e^{-2\kappa|y|} \, \eta_{\mu \nu} \, dx^{\mu} \, dx^{\nu} +
dy^2 \;,
\end{equation}
instead of expression \eqref{metric}. As a result, the hierarchy
relation is given by the formula
\begin{equation}\label{RS_hierarchy_relation_uncorr}
\bar{M}_{\mathrm{Pl}}^2 = \frac{\bar{M}_5^3}{\kappa} \left( 1 -
e^{-2 \pi \kappa r} \right) \;,
\end{equation}
with the graviton masses
\begin{equation}\label{graviton_masses_uncorr}
m_n = x_n \, \kappa \, e^{- \pi \kappa r}, \qquad n=1,2 \ldots \;.
\end{equation}

As for the parameter $\Lambda_{\pi}$, it is defined by
\begin{equation}\label{lambda_uncorr}
\Lambda_{\pi} = \bar{M}_{\mathrm{Pl}} \,  e^{-\pi \kappa r} \;.
\end{equation}

Given $\kappa r \simeq 12$, the scale $\Lambda_{\pi}$
\eqref{lambda_uncorr} is equal to one or several TeV. In order the
hierarchy relation
\eqref{RS_hierarchy_relation_uncorr} to be satisfied, one has to put%
\footnote{Within the bounds $0.01 \leqslant \kappa/\bar{M}_5
\leqslant 0.1$~\cite{Davoudiasl:01}.}
\begin{equation}\label{scale_relation_uncorr}
\kappa \sim \bar{M}_5 \sim \bar{M}_{\mathrm{Pl}} \;.
\end{equation}
Then it follows from \eqref{graviton_masses_uncorr} that the
lightest modes of the KK graviton have masses around 1 TeV.

Thus, one obtains a series of massive graviton resonances in the TeV
region which interact rather strongly with the SM fields, since
$\Lambda_{\pi} \sim 1$ TeV on the visible brane. The experimental
signature of the ``\emph{large curvature option}'' of the RS model
is the real or virtual production of the massive KK graviton
resonances.

The real production of these gravitons could be detected at the
Tevatron in the processes $p \bar{p} \rightarrow \gamma + G^{(n)}$
or $p \bar{p} \rightarrow jet + G^{(n)}$. Narrow high-mass
resonances can be seen in Drell-Yan, di-photon, and di-jet events,
$p \bar{p} \rightarrow G^{(n)} \rightarrow \mu^+ \mu^-, \ \gamma
\gamma, \ 2jets$.

The recent limits on the mass of the lightest KK mode, $m_1$, come
from the Tevatron measurements of $e^+e^-$ and $\gamma \gamma$ final
states~\cite{Tevatron_limits}:
\begin{eqnarray}\label{Tevatron_limits}
m_1 &>& 875 \mathrm{\ MeV}, \qquad (\mathrm{CDF}) \;,
\nonumber\\
m_1 &>& 865 \mathrm{\ MeV}, \qquad  \; (\mathrm{D0}) \;.
\end{eqnarray}
The LHC search limit on $m_1$ is \cite{LHC_limits}
\begin{equation}\label{LHC_dijets}
m_1 \sim 0.7- 0.8 \mathrm{\ TeV}
\end{equation}
for the di-jet mode and integrated luminosity $\mathcal{L}= 0.1
\mathrm{\ fb}^{-1}$, while for the di-photon mode and $\mathcal{L}=
10 \mathrm{\ fb}^{-1}$ the estimate looks like~\cite{LHC_limits}
\begin{equation}\label{LHC_diphoton}
m_1 \sim 1.31 - 3.47 \mathrm{\ TeV} \;.
\end{equation}
Let us underline that these (and analogous) experimental bounds can
not be applied to the RS scheme with the small curvature.

Note, in the case of large curvature \eqref{scale_relation_uncorr}
which arise in the metric \eqref{metric_uncorr}, the size of the
$AdS_5$ slice should be extremely small. Namely,
\begin{equation}\label{radius}
r \simeq 60 \, l_{\mathrm{Pl}} \;,
\end{equation}
where $l_{\mathrm{Pl}}$ is a Planck scale. Thus, in order to
explain the value of $\bar{M}_{\mathrm{Pl}}$ ($\sim 10^{19}$~GeV)
through the fundamental (5-dimensional) Planck scale ($\sim
10^3$~GeV), one has to introduce new huge mass scales $\bar{M}_5$,
$\kappa$, as well as $1/r$. In other words, \emph{hierarchy
problem is not solved, but reformulated} in terms of the new
parameter related to the size of
the bulk along the extra dimension.%
\footnote{A similar shortcoming exists in models with large extra
dimensions of a size $R$, in which a new large mass scale, $1/R$,
is introduced in order to explain the value of $M_{\mathrm{Pl}}$.}

Moreover, there exists another shortcoming of the scenario with the
curvature and fundamental gravity scale being of the order of the
Planck mass. Namely, \emph{kinetic terms} of all graviton fields on
the TeV brane \emph{does not have a canonical form}, and Lorentz
indices are raised with the Minkowski tensor, while the metric in
the coordinates $x^{\mu}$ is $e^{- \pi \kappa r}\eta_{\mu
\nu}$~\cite{Boos:02}.

The correct interpretation of the effective 4-dimensional theory and
correct determination of the masses can be achieved by changing
variables:
\begin{equation}\label{change_variables}
x^{\mu} \rightarrow z^{\mu} =  x^{\mu} e^{- \pi \kappa r} \;.
\end{equation}
As one can see, the metric \eqref{metric_uncorr} turns into the
metric \eqref{metric} under such a replacement.

The ``\emph{small curvature option}'' was studied in the previous
papers~\cite{Kisselev:05,Kisselev:06} (see also
Ref.~\cite{Giudice:04} in which this model was proposed for the
first time). In what follows, the 5-dimensional reduced Planck mass
$\bar{M}_5$ is taken to be one or few TeV. Following
Ref.~\cite{Kisselev:05}-\cite{Giudice:04}, we chose the parameter
$\kappa$ to be $100 \mathrm{\ MeV} - 1\mathrm{\ GeV}$. These values
of $\kappa$ obey the bounds derived in Ref.~\cite{Kisselev:05}:
\begin{equation}\label{curvature_limits}
10^{-5} \leqslant \frac{\kappa}{\bar{M}_5} \leqslant 0.1 \;.
\end{equation}
Then the mass of the lightest KK excitation lies within the limits
$m_1 \simeq 0.38 - 3.8$~GeV.

Correspondingly, the mass scale $\Lambda_{\pi}$~\eqref{lambda} is
given by
\begin{equation}\label{lambda_enum}
\Lambda_{\pi} = 100 \left( \frac{M_5}{\mathrm{TeV}} \right)^{3/2}
\left( \frac{\mathrm{100 \ MeV}}{\kappa} \right)^{1/2} \!
\mathrm{TeV} \;.
\end{equation}
It immediately follows from Eqs.~\eqref{Lagrangian},
\eqref{lambda_enum} that there is no problem with the radion field
$\phi$ in our scheme, since its coupling to the SM fields ($\sim
1/\sqrt{3}\Lambda_{\pi}$) is strongly suppressed.

As for the massive gravitons, their couplings are also defined by
the scale $\Lambda_{\pi}$~\eqref{lambda_enum}. However, the
smallness of the coupling is compensated by the large number of the
gravitons that can be produced in any inclusive process. As a
result, magnitudes of corresponding cross sections will be defined
by the 5-dimensional gravity scale $\bar{M}_5$ (see our comments
after Eq.~\eqref{parameter_A_epsilon}).

Thus, we have an infinite number of low-mass KK resonances with the
small mass splitting, in contrast with the usually adopted RS
scenario~\eqref{scale_relation_uncorr}. Nevertheless, due to the
warp geometry of the $AdS_5$ space-time, the RS model with the small
curvature differs significantly from the ADD
model~\cite{Arkani-Hamed:98} (at least at $\kappa \gg 10^{-22}$ eV),
as was demonstrated in Ref.~\cite{Kisselev:06}.

In Ref.~\cite{Kisselev:05} this scheme was applied for the elastic
scattering of the brane fields at ultra-high energies induced by
$t$-channel gravireggeons~\cite{Kisselev:04}.

Recently, the small curvature option of the RS model has been
checked experimentally by the DELPHI Collaboration. The gravity
effects were searched for by studying photon energy spectrum in the
process $e^+e^- \rightarrow \gamma + E_{\perp}\hspace{-6mm} \diagup
\hspace{2mm}$. No deviations from the SM prediction were seen. As a
result, the following (\emph{preliminary}) bound was
obtained~\cite{LEP_limit}:
\begin{equation}\label{n=1_limit}
M_5 > 1.69 \mathrm{\ TeV} \pm 3 \% \;,
\end{equation}
that correspond to the reduced 5-dimensional scale $\bar{M}_5 >
0.92$ TeV (see the relation between $M_5$ and $\bar{M}_5$ after
Eq.~\eqref{RS_hierarchy_relation}).

Note that this limit could not be inferred from the limits already
given for larger than two flat dimensions owing to the totally
different spectrum of the photon~\cite{LEP_limit}.

\section{Gravity effects in $\mathbf{e^+e^-}$ annihilation resulting from virtual
KK gravitons}
\label{sec:virtual_gravitons}

Let us now consider the process of $e^+e^-$ annihilation into two
leptons mediated by massive graviton exchanges,
\begin{equation}\label{process}
e^+ \, e^- \rightarrow G^{(n)} \rightarrow l \, \bar{l} \; ,
\end{equation}
where $l = e$, or $\mu$.%
\footnote {The processes $e^+e^- \rightarrow \tau^+ \tau^-, \ \gamma
\gamma$ are also promising reactions at the LC. We will not consider
them in the present paper.}

In what follows, the collision energy, $\sqrt{s}$, is taken to be
equal to 1 TeV (for comparison, $\sqrt{s} = 200$ GeV will be also
considered). It means that we are working in the following region:
\begin{equation}\label{energy_region}
 \Lambda_{\pi} \gg \sqrt{s} \sim M_5 \gg \kappa.
\end{equation}

The matrix element of the process~\eqref{process} looks like
\begin{equation}\label{matrix_element}
\mathcal{M} = \mathcal{A} \, \mathcal{S} \; .
\end{equation}
The fist factor in Eq.~\eqref{matrix_element} contains the following
contraction of tensors:
\begin{equation}\label{tensor_contraction}
\mathcal{A} = T_{\mu \nu}^e \, P^{\mu \nu \alpha \beta} \, T_{\alpha
\beta}^l = T_{\mu \nu}^e \, T^{l \, \mu \nu} - \frac{1}{3} \,
{(T^e)}^{\mu}_{\mu} \, {(T^l)}^{\nu}_{\nu} \; ,
\end{equation}
where $P^{\mu \nu \alpha \beta}$ is the tensor part of the graviton
propagator, while $T_{\mu \nu}^{e}$ ($T_{\mu \nu}^{l}$) is the
lepton energy-momentum tensor.

The second factor in Eq.~\eqref{matrix_element} is \emph{universal}
for all types of processes mediated by $s$ or $t$-channel
exchange of the massive  KK excitations.%
\footnote{For the $\mu^+\mu^-$ final state, there is no $t$-channel
contributions, while in the Bhabha scattering both $\mathcal{S}(s)$
and $\mathcal{S}(t)$ contribute.}
For instance, the graviton exchange in the $s$-channel leads to the
expression
\begin{equation}\label{KK_sum}
\mathcal{S}(s) =  \frac{1}{\Lambda_{\pi}^2} \sum_{n=1}^{\infty}
\frac{1}{s - m_n^2 + i \, m_n \Gamma_n} \; .
\end{equation}
Here $\Gamma_n$ denotes the total width of the graviton with the KK
number $n$ and mass $m_n$. The width is small if $n$ is not too
large~\cite{Kisselev:05_2}:
\begin{equation}\label{graviton_widths}
\frac{\Gamma_n}{m_n} = \eta \left( \frac{m_n}{\Lambda_{\pi}}
\right)^2 ,
\end{equation}
where $\eta \simeq 0.09$.

Note, however, that the main contribution to sum~\eqref{KK_sum}
comes from the region $n \sim \sqrt{s}/\kappa \gg 1$. So,\emph{
nonzero widths of the gravitons} in the RS model with the small
curvature \emph{should be taken into account}.

The sum in Eq.~\eqref{KK_sum} can be calculated analytically by the
use of the formula~\cite{Watson}
\begin{equation}\label{zeros_Bessel_sum}
\sum_{n=1}^{\infty} \frac{1}{ z_{n, \nu}^2 - z^2} = \frac{1}{2 z}
\, \frac{J_{\nu + 1}(z)}{J_{\nu}(z)} \; ,
\end{equation}
where $z_{n, \, \nu}$ ($n=1,2 \ldots$) are zeros of the function
$z^{-\nu} J_{\nu}(z)$. As a result, the following explicit
expression was obtained in Ref.~\cite{Kisselev:06}:
\begin{equation}\label{main_formula}
\mathcal{S}(s) = - \frac{1}{2\kappa \bar{M}_5^3} \, \frac{1}{\sqrt{1
- \displaystyle 4 i \, \eta \frac{s}{\Lambda_{\pi}^2}}} \left[
\frac{1}{\sigma} \, \frac{J_2(\sigma)}{J_1(\sigma)} - \frac{1}{\rho}
\, \frac{J_2(\rho)}{J_1(\rho)} \right] \;,
\end{equation}
where
\begin{align}
\sigma &= \frac{\sqrt{s}}{\kappa} \Bigg[ \frac{2}{1 + \sqrt{1 -
\displaystyle  4 i \eta \frac{s}{\Lambda_{\pi}^2}}} \Bigg]^{1/2}
\label{sigma} \\
\intertext{and} \rho &= \frac{1}{\sqrt{2i \eta}}
\frac{\Lambda_{\pi}}{\kappa} \left[ 1 + \sqrt{1 - 4 i \eta
\frac{s}{\Lambda_{\pi}^2}} \right]^{1/2} \;. \label{rho}
\end{align}

Since $\eta s/\Lambda_{\pi}^2 \ll 1$ in our case, we obtain from
\eqref{sigma}, \eqref{rho}:
\begin{align}
\sigma & \simeq \frac{\sqrt{s}}{\kappa} + \frac{i \eta}{2} \, \left(
\frac{\sqrt{s}}{\bar{M}_5}\right)^3
\label{sigma_appr} \\
\rho & \simeq \frac{1}{\sqrt{\eta}} \, \frac{\Lambda_{\pi}}{\kappa}
\gg |\,\sigma| \;, \label{rho_appr}
\end{align}
and Eq.~\eqref{main_formula} becomes~\cite{Kisselev:06}
\begin{equation}\label{main_formula_asym}
\mathcal{S}(s) = - \frac{1}{2 \kappa \bar{M}_5^3} \,
\frac{1}{\sigma} \, \frac{J_2(\sigma)}{J_1(\sigma)} \; ,
\end{equation}
with $\sigma$ given by \eqref{sigma_appr} (here and in what follows,
small corrections $\mathrm{O(\kappa/\sqrt{s})}$ are omitted).

By using the asymptotics of the Bessel functions~\cite{Watson} the
formula \eqref{main_formula} can be represented in the final
form~\cite{Kisselev:06}:
\begin{equation}\label{main_formula_final}
\mathcal{S}(s) = - \frac{1}{4 \bar{M}_5^3 \sqrt{s}} \; \frac{\sin 2A
+ i \sinh 2\varepsilon }{\cos^2 \! A + \sinh^2 \! \varepsilon } \; ,
\end{equation}
where
\begin{equation}\label{parameter_A_epsilon}
A = \frac{\sqrt{s}}{\kappa} + \frac{\pi}{4}, \qquad \varepsilon  =
\frac{\eta}{2} \Big( \frac{\sqrt{s}}{\bar{M}_5} \Big)^3 \; .
\end{equation}

As one can see from \eqref{main_formula_final}, the magnitude of
$\mathcal{S}(s)$ is defined by $\bar{M}_5$ and $\sqrt{s}$, not by
$\Lambda_{\pi}$, although the latter describes the graviton coupling
with the matter in the effective Lagrangian \eqref{Lagrangian}.

The following inequalities immediately result from
\eqref{main_formula_final}:
\begin{equation}\label{bounds_ImS}
- \coth \varepsilon  \leqslant \mathrm{Im}
\,\mathcal{\tilde{S}}(s) \leqslant - \tanh \varepsilon  \;
\end{equation}
\begin{align}
\big| \mathrm{Re} \,\mathcal{\tilde{S}}(s) \big| &\leqslant
\frac{1}{1 + 2\sinh^2
\! \varepsilon } \; ,
\label{bounds_ReS}  \\
\left| \frac{\mathrm{Re} \, \mathcal{\tilde{S}}(s)}{\mathrm{Im}
\,\mathcal{\tilde{S}}(s)} \right| &\leqslant \frac{1}{\sinh 2
\varepsilon } \;,
\label{bound_ReS/ImS}
\end{align}
where the notation $ \mathcal{\tilde{S}}(s) = [2 \bar{M}_5^3
\sqrt{s} \,] \, \mathcal{S}(s)$ was introduced. Note that the upper
bound for the ratio $|\, \mathrm{Re} \, \mathcal{S}(s)/\mathrm{Im}
\,\mathcal{S}(s)|$ decreases with energy, and it becomes as small as
0.08 at $\sqrt{s} \simeq 3 \bar{M}_5$. At the same time,
$\mathrm{Im} \,\mathcal{S}(s)$ tends to the value $-1/(2 \bar{M}_5^3
\sqrt{s})$.

Should one ignores the widths of the massive gravitons, and then
replace a summation in KK number \eqref{KK_sum} by integration over
graviton masses,%
\footnote{By using the relation $dn = \bar{M}_{\mathrm{Pl}}^2/(2
\pi \bar{M}_5^3) \; dm$.}
one gets (see, for instance, \cite{Kisselev:06,Giudice:04}):
\begin{equation}\label{KK_sum_zero_widths}
\mathrm{Im} \,\mathcal{S}(s) = - \frac{1}{2 \bar{M}_5^3 \sqrt{s}}
\;, \qquad \mathrm{Re} \, \mathcal{S}(s) = 0 \;,
\end{equation}
in contrast to the exact formula \eqref{main_formula_final}.

However, the series of low-massive resonances in the RS model with
the small curvature can be replace by the continuous spectrum
\emph{only in a trans-Planckian energy region}  $\sqrt{s} \gtrsim 3
\bar{M}_5$. It can be understood as follows. One may regard the set
of narrow graviton resonances to be the continuous mass spectrum
(within a relevant interval of $n$), if only
\begin{equation}\label{continuous_spectrum}
\Delta m_{\mathrm{KK}} < \Gamma_n
\end{equation}
is satisfied, where $\Delta m_{\mathrm{KK}}$ is the mass splitting.
As was shown in Ref.~\cite{Kisselev:06}, the inequality
\eqref{continuous_spectrum} is equivalent to the above mentioned
inequality
\begin{equation}\label{trans_planckian_region}
\sqrt{s} \gtrsim 3 \bar{M}_5 \; .
\end{equation}

We are working in another kinematical region, $\sqrt{s} \lesssim 3
\bar{M}_5$, since the 5-dimensional scale $\tilde{M}_5$ is assumed
to be equal to (or larger than) 1 TeV, while the collision energy is
fixed to be 1 TeV. That is why, for our calculations we will use the
analytical expression~\eqref{main_formula_final} which takes into
account both the discrete character of the graviton spectrum and
finite widths of the KK gravitons.

Let us note that our formula~\eqref{main_formula_final} can be also
applied to the scattering of the brane particles, induced by
exchanges of $t$-channel gravitons~\cite{Kisselev:05,Kisselev:05_2}.
In such a case, it looks like
\begin{equation}\label{t_channel}
\mathcal{S}(t) = - \frac{1}{2 \kappa \bar{M}_5^3} \,
\frac{1}{\tilde{\sigma}} \,
\frac{I_2(\tilde{\sigma})}{I_1(\tilde{\sigma})} \; ,
\end{equation}
where $I_{\nu}(z) = \exp (-i\nu \pi/2) \, J_{\nu} (i z)$ is the
modified Bessel function, and
\begin{equation}\label{sigma_tilde}
\tilde{\sigma} \simeq  \frac{\sqrt{-t}}{\kappa} - \frac{i \eta}{2}
\, \left( \frac{\sqrt{-t}}{\bar{M}_5}\right)^3 \;,
\end{equation}
with $t$ being a 4-momentum transfer. Since $I_2(z)/I_1(z)
\rightarrow 1$ at $z \gg 1$ ($- \pi/2 < \arg z < 3
\pi/2$)~\cite{Watson}, we find that $\mathcal{S}(t)$ is pure real in
the kinematical region $\bar{M}_5^3/\kappa \gg - t \gg
\kappa^2$~\cite{Kisselev:06}:
\begin{equation}\label{t_channel_final}
\mathcal{S}(t) = - \frac{1}{2\bar{M}_5^3 \sqrt{-t}} \; .
\end{equation}

Our main goal is to apply theoretical expressions
\eqref{main_formula_final}, \eqref{t_channel_final} for estimating
virtual graviton contributions to the process \eqref{process}. The
relations of cross sections with the quantities $\mathcal{S}(s)$ and
$\mathcal{S}(t)$ are taken from the Appendix of
Ref.~\cite{Giudice:04}.

\begin{figure}[ht]
\begin{center}
\resizebox{.6 \textwidth}{!}{\includegraphics{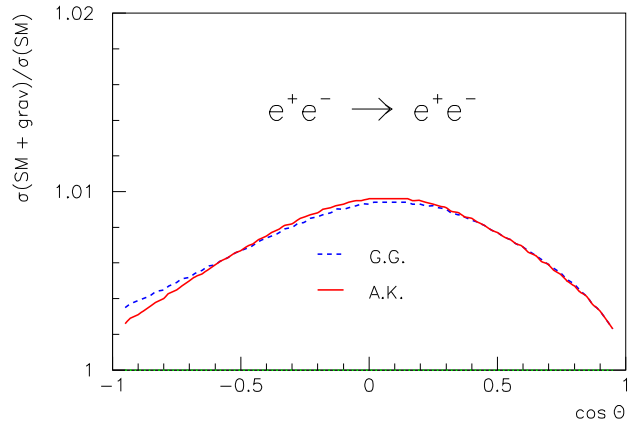}}
\caption{The correction to the Bhabha cross section resulting from
virtual graviton exchanges as a function of the scattering angle at
the LEP2 energy $\sqrt{s} = 200$ GeV.  The solid line is our
prediction which takes into account both a discrete character of the
spectrum and widths of the KK gravitons. The dashed line corresponds
to the continuous mass approximation for the graviton
spectrum~\cite{Giudice:04}. The curves correspond to $M_5 = 1.8$ TeV
and $\kappa = 100$ MeV.} \label{fig:LEP_1.8TeV_0.1GeV}
\end{center}
\end{figure}

Let us consider the Bhabha scattering first. In
Fig.~\ref{fig:LEP_1.8TeV_0.1GeV} the graviton contribution is
presented as a function of the scattering angle of the final leptons
at the collision energy $\sqrt{s} = 200$ GeV (solid line). Another
prediction is also shown (dashed line) which was calculated under
assumption that the dense spectrum of the KK gravitons can be
approximated by the continuum.

We see that the difference between two predictions is negligible at
LEP2 energy. Moreover, the gravity effects are very small with
respect to the SM cross section (see
Fig.~\ref{fig:LEP_1.8TeV_0.1GeV}). Thus, we can conclude that the
above mentioned bounds on $\bar{M}_5$ from LEP~\eqref{n=1_limit}
should be also applied to the 5-dimensional scale in our scheme.

However, these two estimations (based on exact formula
~\eqref{main_formula_final} and approximate
one~\eqref{KK_sum_zero_widths}, respectively) differ drastically
when we consider a collision energy relevant for future linear
colliders~\cite{LC_projects}. This is illustrated in
Figs.~\ref{fig:bhabha_1.8TeV_0.1GeV}--\ref{fig:bhabha_1.8TeV_0.994GeV}.
If we compare Figs.~\ref{fig:bhabha_1.8TeV_0.9GeV},
\ref{fig:bhabha_1.8TeV_0.994GeV} and \ref{fig:bhabha_1.8TeV_1GeV},
we conclude that at fixed value of $M_5$, the ratio of the total
cross section to the SM cross section is very sensitive to
variations of the curvature $\kappa$ within a narrow region (0.9 GeV
-- 1 GeV, in our case). The effect is less pronounced, but still
significant, at larger $M_5$ (see
Figs.~\ref{fig:bhabha_2.5TeV_0.1GeV}, \ref{fig:bhabha_2.5TeV_1GeV}).
Let us note that the interference of gravity with the SM forces is
constructive in the Bhabha scattering, and the ratio
$\sigma(\mathrm{SM+grav})/\sigma(\mathrm{SM})$ is lager than 1 for
all cases.

Now let us consider the process $e^+e^- \rightarrow \mu^+ \mu^-$.
The estimates show that the gravity effects are negligible at
$\sqrt{s} = 200$ GeV. Contrary to the Bhabha process, the gravity
effects calculated with the use of approximate
formulas~\eqref{KK_sum_zero_widths} become very small at $M_5 = 2.5$
TeV (see
Figs.~\ref{fig:muons_1.8TeV_0.1GeV}--\ref{fig:muons_2.5TeV_1GeV}).
As for predictions based on the exact formula
\eqref{main_formula_final}, they change both qualitatively and
quantitatively with variations of the parameters of the RS model. To
see this, it is enough to compare
Figs.~\ref{fig:muons_1.8TeV_0.1GeV} and \ref{fig:muons_1.8TeV_1GeV}
($\kappa$ changes from 100 MeV to 1 GeV, with $M_5$ being fixed), as
well as Figs.~\ref{fig:muons_1.8TeV_1GeV} and
\ref{fig:muons_2.5TeV_1GeV} ($\kappa$ is fixed, while $M_5$ changes
from 1.8 TeV to 2.5 TeV). The interference can be destructive or
constructive, depending on $\cos \theta$ and values of the
parameters.

%%%%%%%%%%%%%%%
% Conclusions %
%%%%%%%%%%%%%%%

\section{Conclusions and discussions}

In the present paper the contributions of the virtual $s$ and
$t$-channel KK gravitons to the Bhabha scattering and to the process
$e^+ e^- \rightarrow \mu^+ \mu^-$ at the LC energy $\sqrt{s} = 1$
TeV were numerically estimated. We have considered the small
curvature option of the RS model with two branes ($\kappa \ll
\bar{M}_5$). In such a scheme, the KK graviton spectrum is the
series of the narrow low-mass resonances. The SM fields live on the
TeV brane, while the gravity propagates in the bulk.

The 5-di\-mensional Planck scale $M_5$ is taken to be 1.8 TeV and
2.5 TeV, with the curvature parameter being restricted to the region
$\kappa = 100 \mathrm{\ MeV} \div 1 \mathrm{\ GeV}$, that means
$\sqrt{s} \sim M_5 \gg \kappa$. For the numerical estimates, we used
the formula for the process-independent gravity part of the
scattering amplitude, $\mathcal{S}(s)$, from Ref.~\cite{Kisselev:06}
(see Eq.~\eqref{main_formula_final}).

For comparison, we have made calculations by using
Eqs.~\eqref{KK_sum_zero_widths} which treat the spectrum of the KK
graviton as the continuum. We have found that both expressions
coincide only at the LEP2 energy, when the gravity effects appeared
to be small. At the LC energy their predictions differ both
qualitatively and quantitatively. It means that the sum in the KK
number can not be approximated by integration over graviton mass.
Moreover, the graviton widths should be properly taken into account,
as formula \eqref{main_formula_final} does.

\begin{figure}[h]
\begin{center}
\resizebox{.6 \textwidth}{!}{\includegraphics{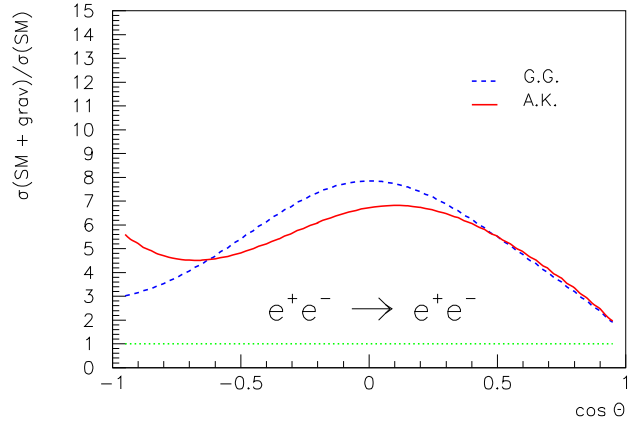}}
\caption{The same as in Fig.~\ref{fig:LEP_1.8TeV_0.1GeV}, except
for the collision energy is equal to $\sqrt{s} = 1$ TeV.}
\label{fig:bhabha_1.8TeV_0.1GeV}
\end{center}
\end{figure}

\begin{figure}[h]
\begin{center}
\resizebox{.6 \textwidth}{!}{\includegraphics{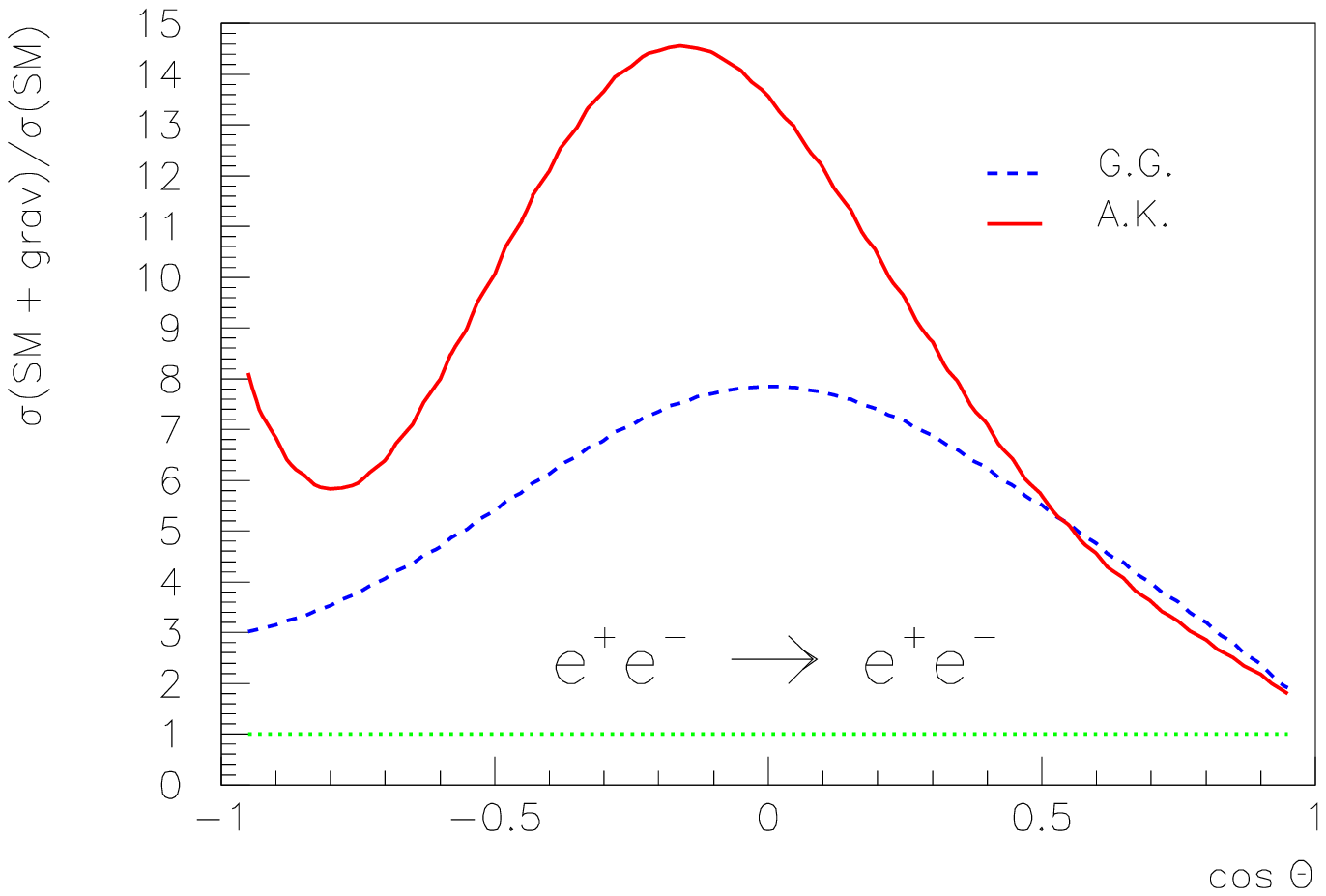}}
\caption{The same energy as in
Fig.~\ref{fig:bhabha_1.8TeV_0.1GeV}, with the parameters $M_5 =
1.8$ TeV, $\kappa = 1$ GeV.} \label{fig:bhabha_1.8TeV_1GeV}
\end{center}
\end{figure}

\begin{figure}[h]
\begin{center}
\resizebox{.6 \textwidth}{!}{\includegraphics{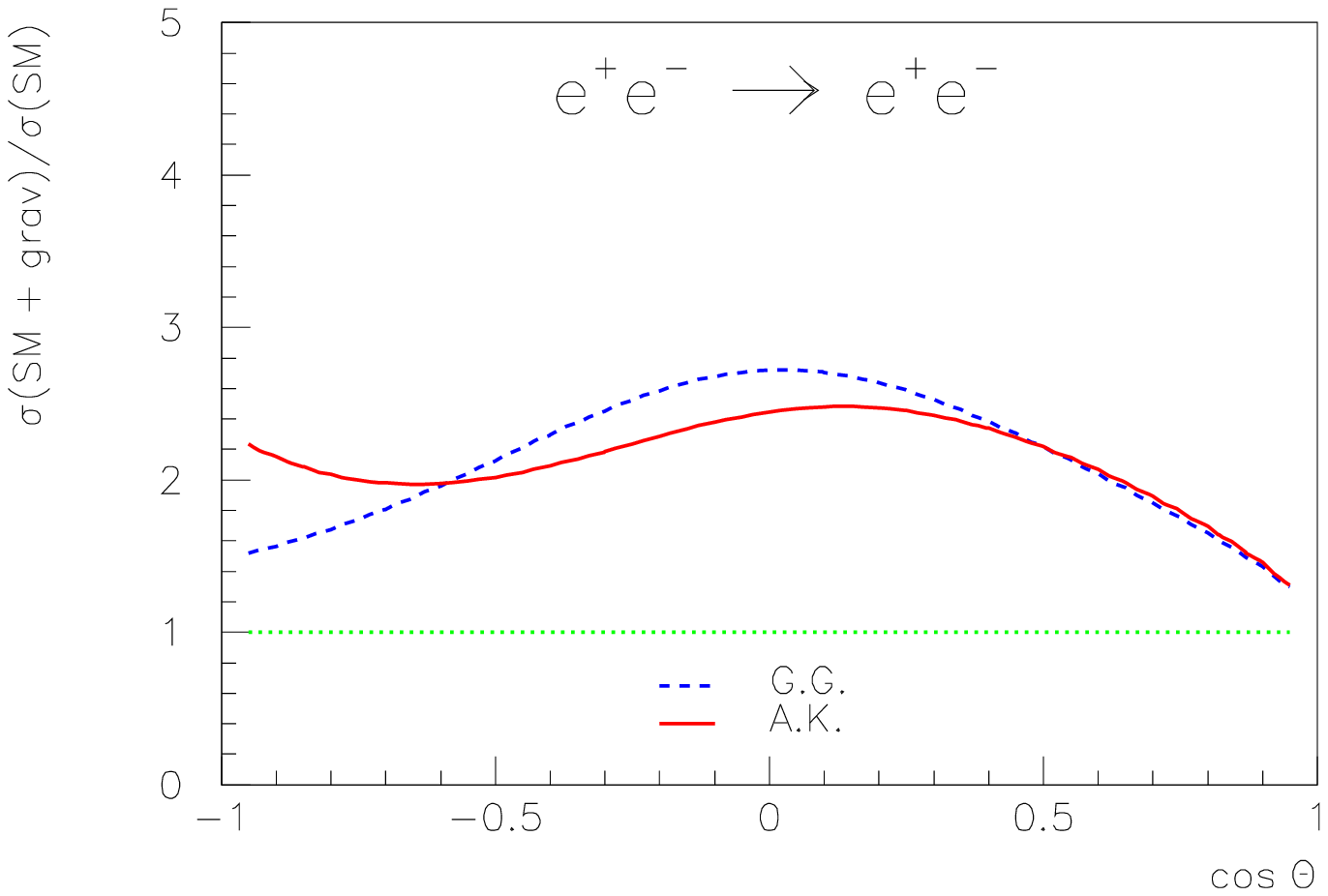}}
\caption{The same energy as in
Fig.~\ref{fig:bhabha_1.8TeV_0.1GeV}, with the parameters $M_5 =
2.5$ TeV, $\kappa = 100$ MeV.} \label{fig:bhabha_2.5TeV_0.1GeV}
\end{center}
\end{figure}

\begin{figure}[h]
\begin{center}
\resizebox{.6 \textwidth}{!}{\includegraphics{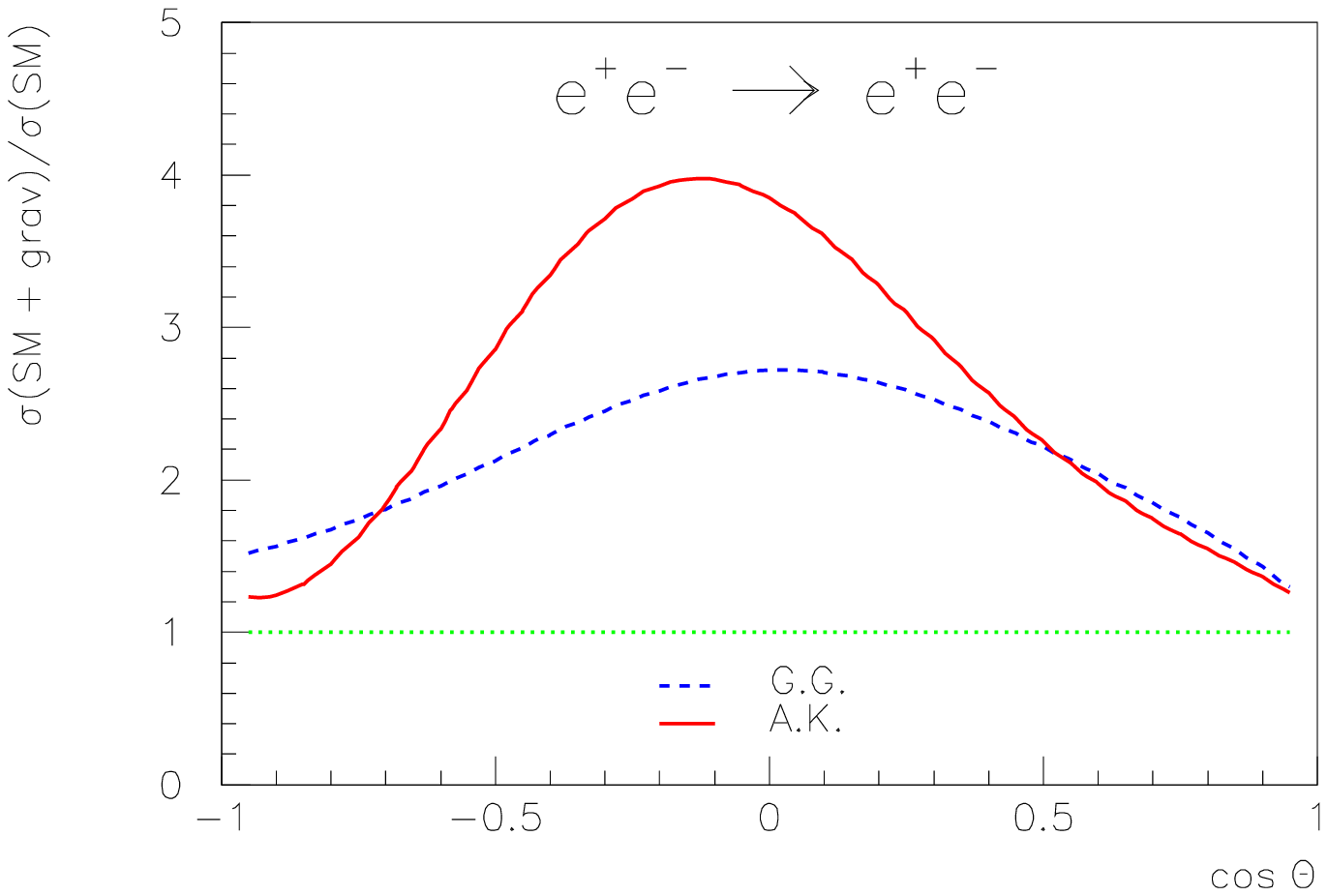}}
\caption{The same energy as in
Fig.~\ref{fig:bhabha_1.8TeV_0.1GeV}, with the parameters $M_5 =
2.5$ TeV, $\kappa = 1$ GeV.} \label{fig:bhabha_2.5TeV_1GeV}
\end{center}
\end{figure}

\begin{figure}[h]
\begin{center}
\resizebox{.6 \textwidth}{!}{\includegraphics{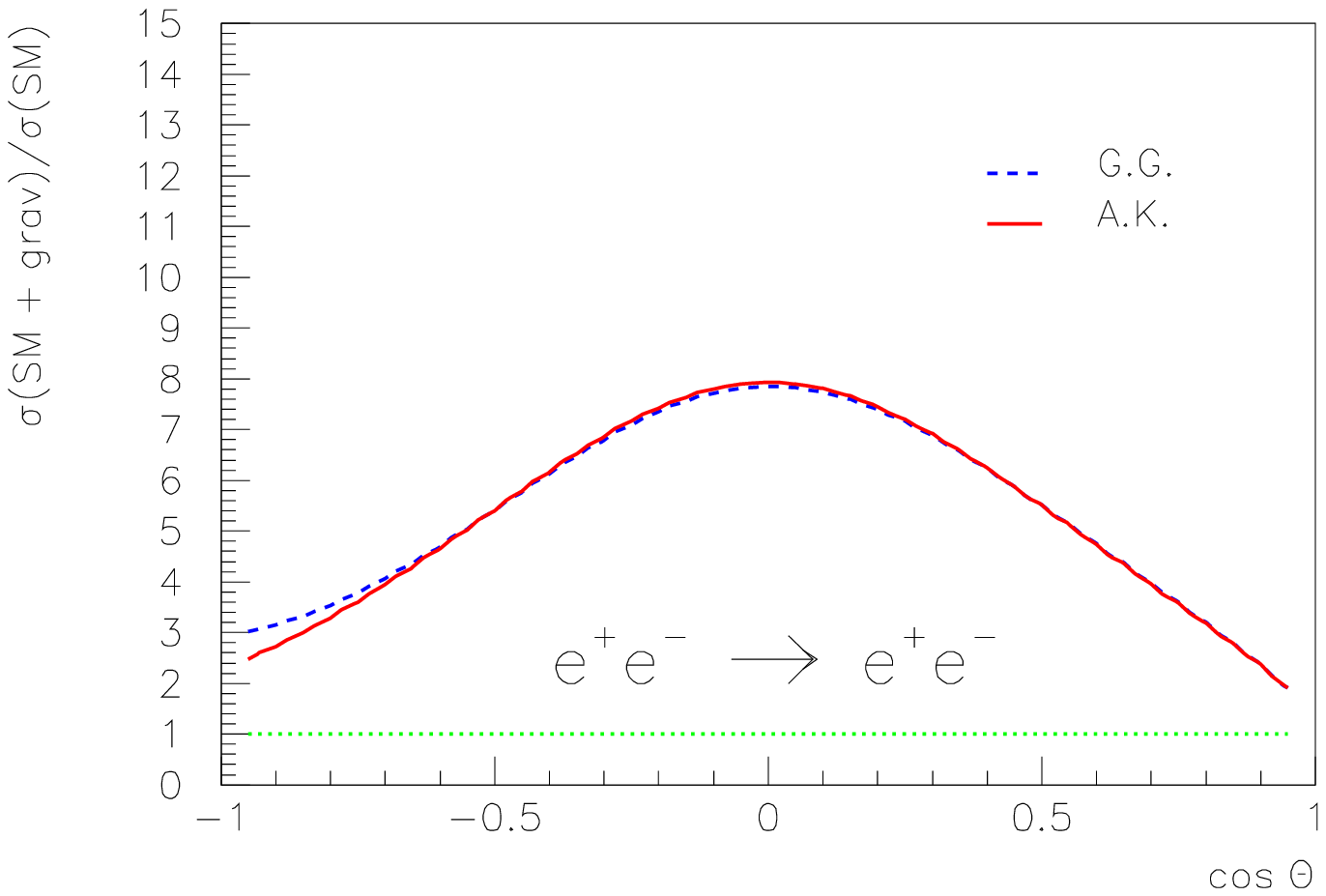}}
\caption{The same energy as in
Fig.~\ref{fig:bhabha_1.8TeV_0.1GeV}, with the parameters $M_5 =
1.8$ TeV, $\kappa = 900$ MeV.} \label{fig:bhabha_1.8TeV_0.9GeV}
\end{center}
\end{figure}

\begin{figure}[h]
\begin{center}
\resizebox{.6 \textwidth}{!}{\includegraphics{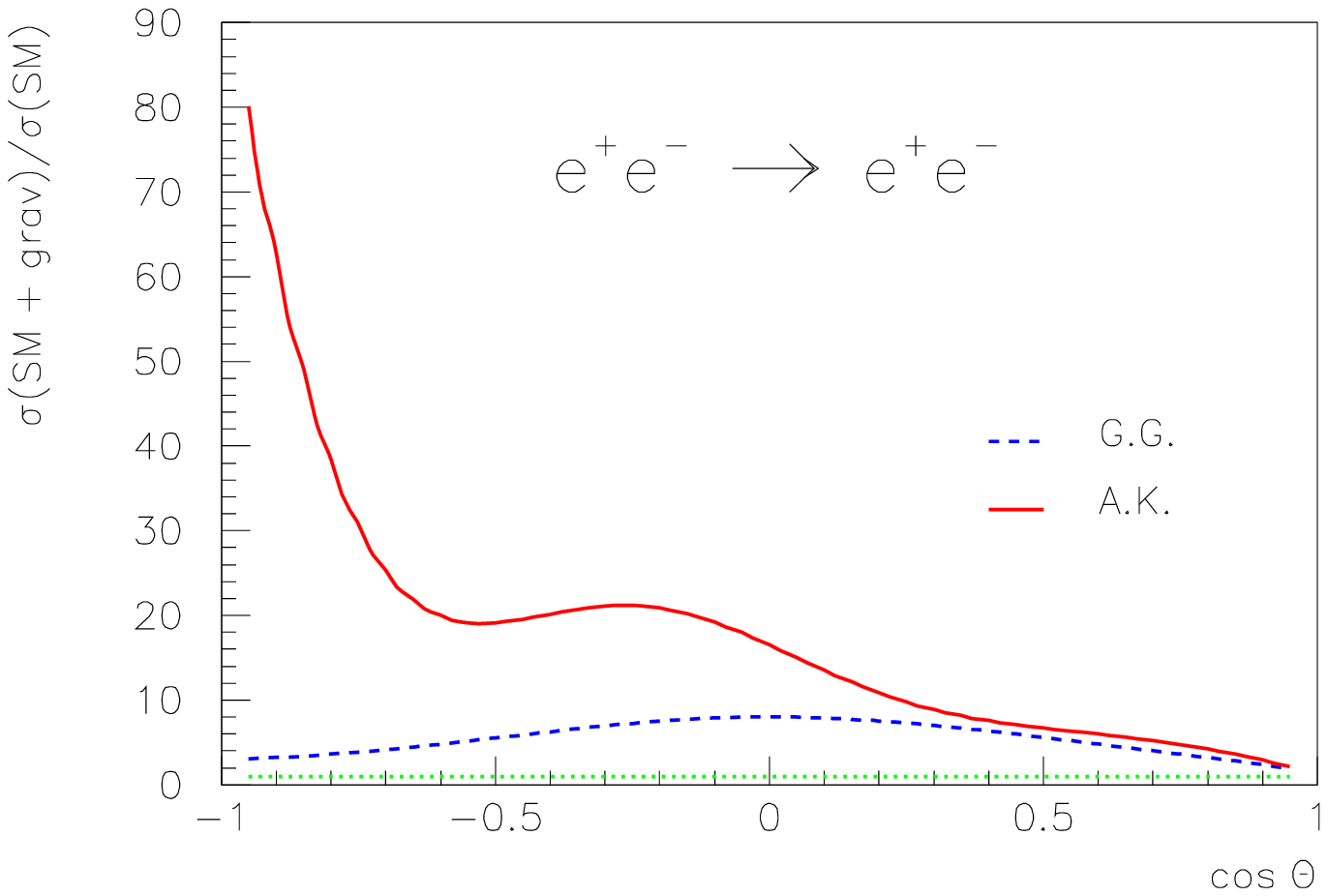}}
\caption{The same energy as in
Fig.~\ref{fig:bhabha_1.8TeV_0.1GeV}, with the parameters $M_5 =
1.8$ TeV, $\kappa = 994$ MeV.} \label{fig:bhabha_1.8TeV_0.994GeV}
\end{center}
\end{figure}

In
Figs.~\ref{fig:bhabha_1.8TeV_0.1GeV}--\ref{fig:bhabha_1.8TeV_0.994GeV}
the ratio $\sigma(\mathrm{SM+grav})/\sigma(\mathrm{SM})$ for the
Bhabha process at collision energy $\sqrt{s} = 1$ TeV is presented.
Everywhere $\sigma(\mathrm{SM})$ means the SM differential cross
section with respect to $\cos \theta$, while
$\sigma(\mathrm{SM+grav})$ means the differential cross section
which takes into account both the SM and gravity interactions.

\begin{figure}[h]
\begin{center}
\resizebox{.6 \textwidth}{!}{\includegraphics{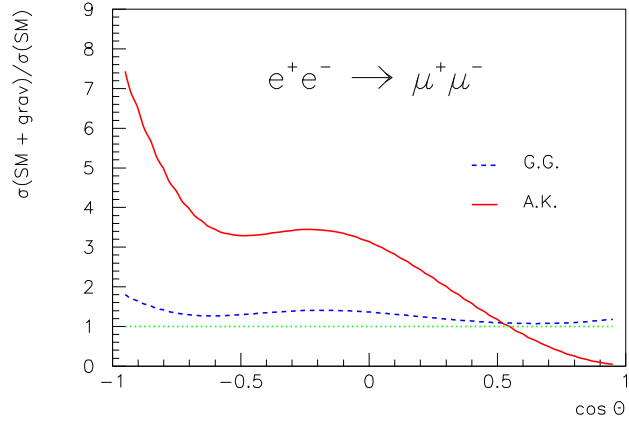}}
\caption{The virtual graviton contribution to the cross section of
the process $e^+e^- \rightarrow \mu^+ \mu^-$ as a function of the
scattering angle. The collision energy $\sqrt{s} = 1$ TeV. Both
curves correspond to $M_5 = 1.8$ TeV, $\kappa = 100$ MeV.}
\label{fig:muons_1.8TeV_0.1GeV}
\end{center}
\end{figure}

\begin{figure}[h]
\begin{center}
\resizebox{.6 \textwidth}{!}{\includegraphics{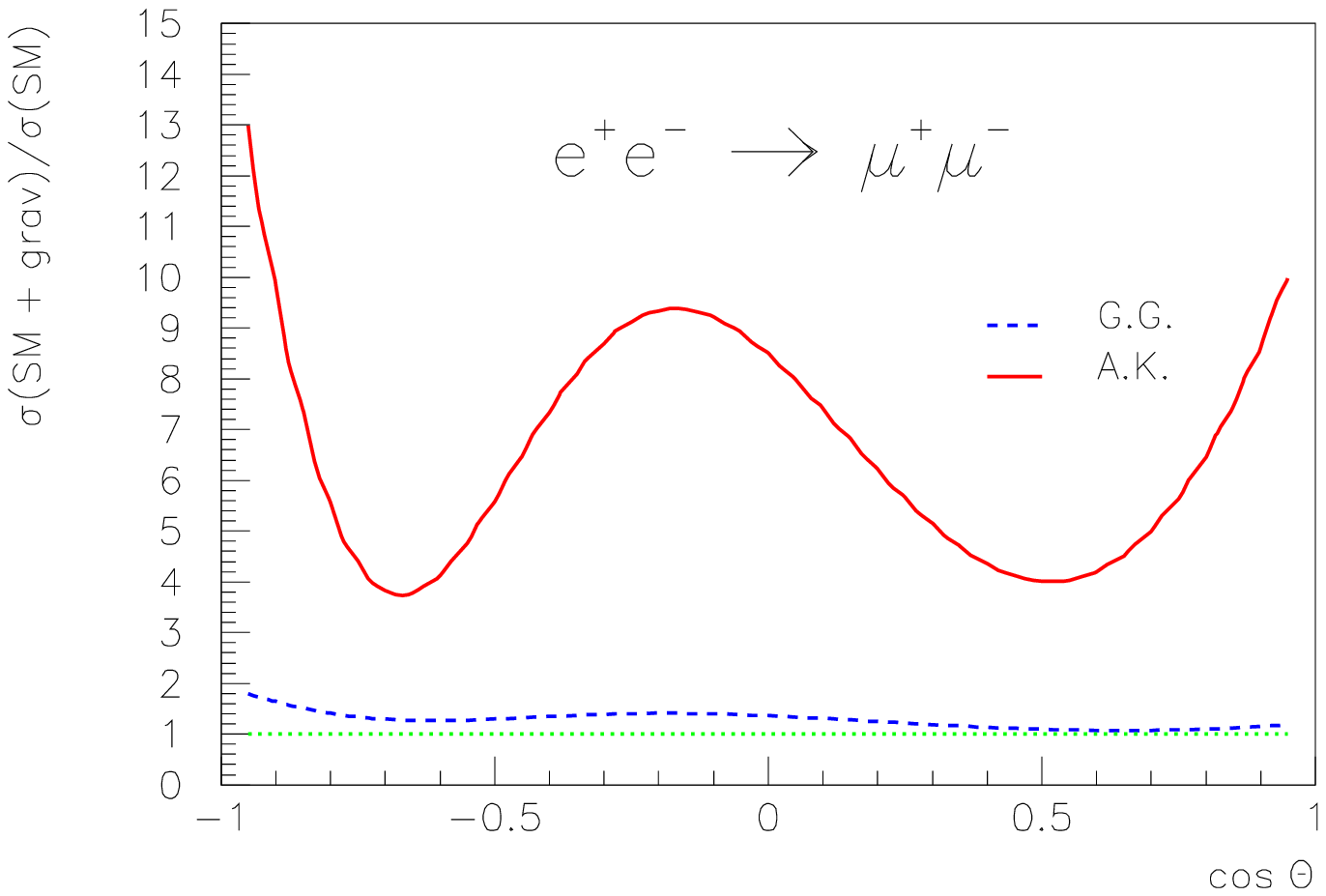}}
\caption{The same as in Fig.~\ref{fig:muons_1.8TeV_0.1GeV}, with
the parameters $M_5 = 1.8$ TeV, $\kappa = 1$ GeV.}
\label{fig:muons_1.8TeV_1GeV}
\end{center}
\end{figure}

\begin{figure}[h]
\begin{center}
\resizebox{.6 \textwidth}{!}{\includegraphics{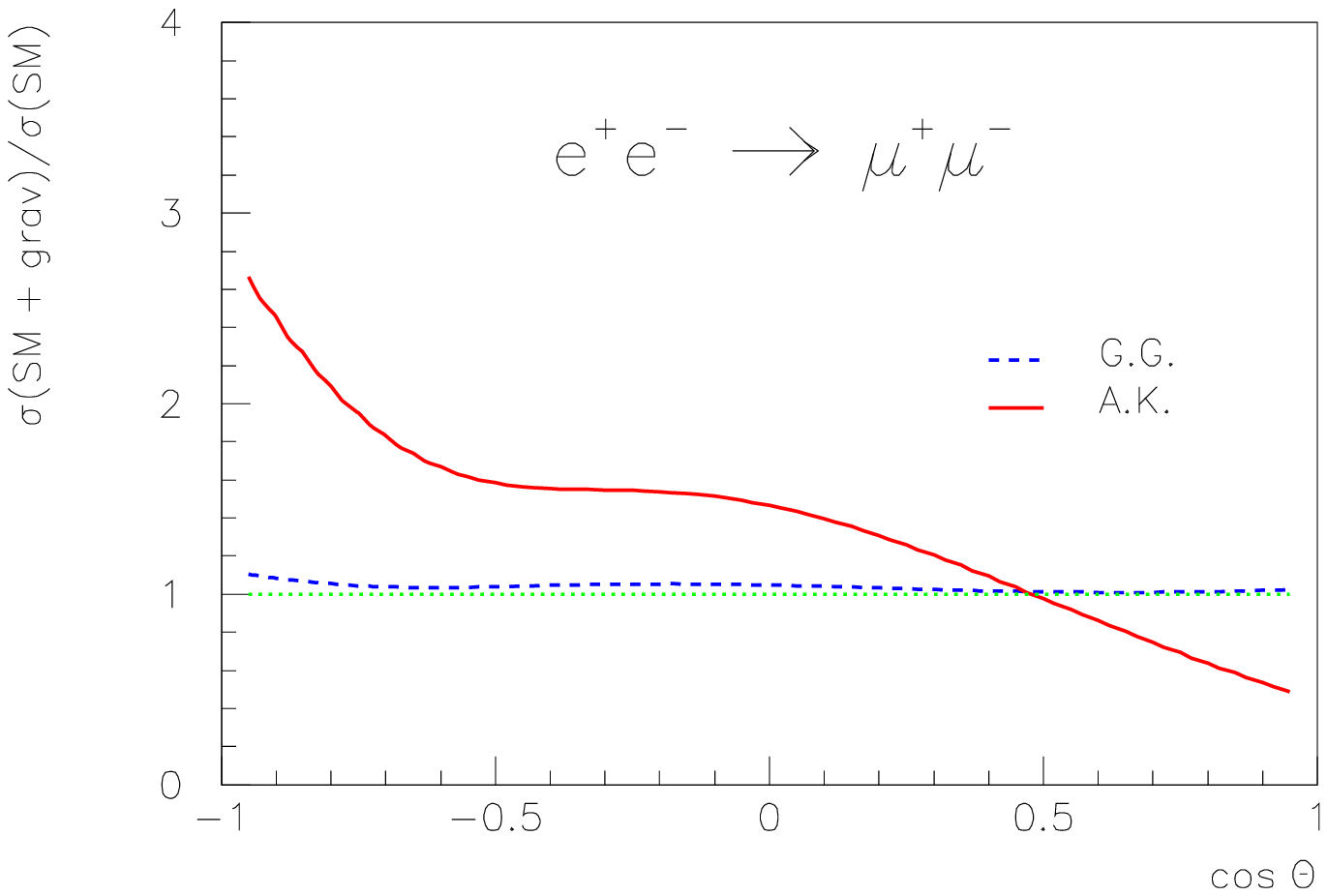}}
\caption{The same as in Fig.~\ref{fig:muons_1.8TeV_0.1GeV}, with
the parameters $M_5 = 2.5$ TeV, $\kappa = 100$ GeV.}
\label{fig:muons_2.5TeV_0.1GeV}
\end{center}
\end{figure}

\begin{figure}[h]
\begin{center}
\resizebox{.6 \textwidth}{!}{\includegraphics{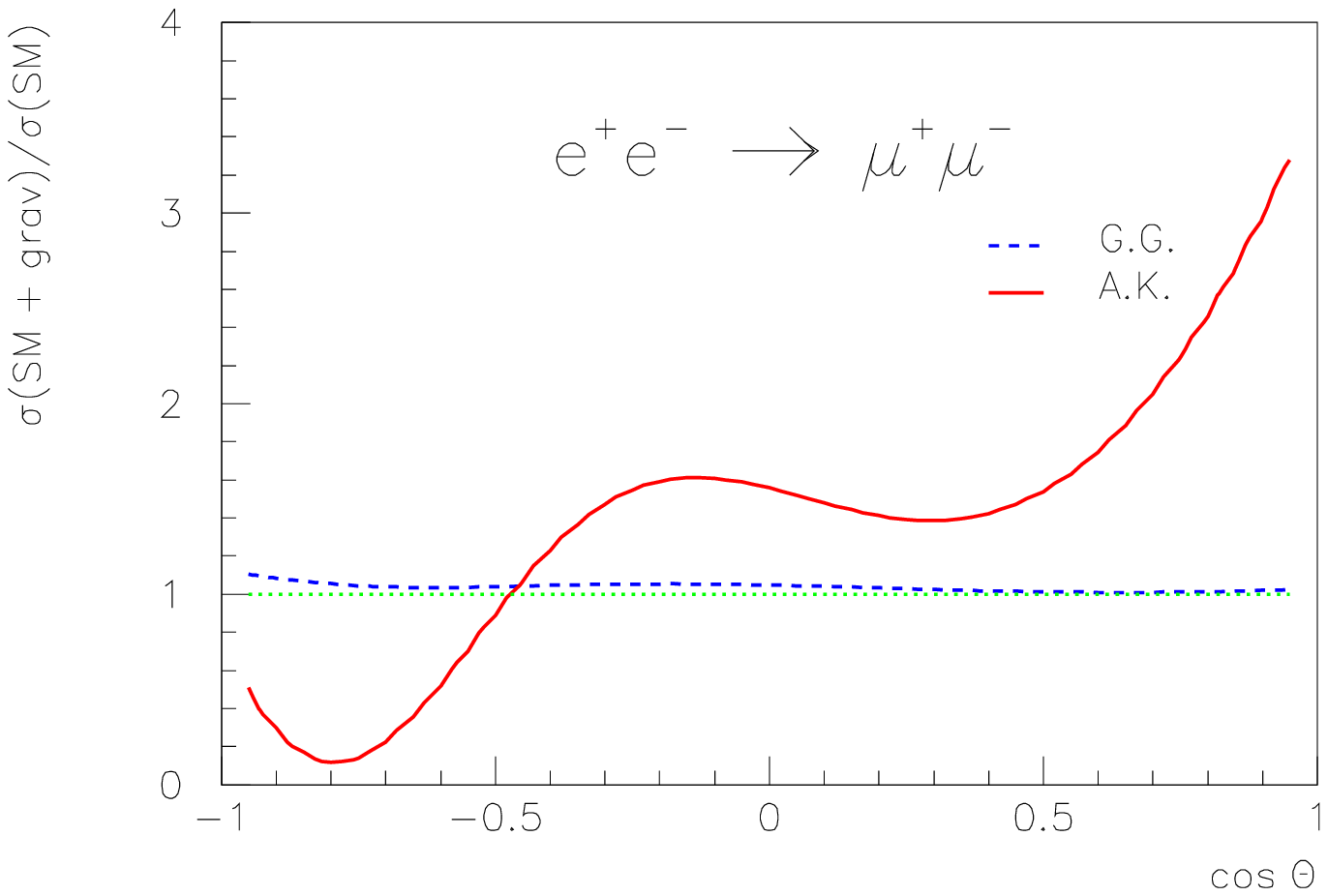}}
\caption{The same as in Fig.~\ref{fig:muons_1.8TeV_0.1GeV}, with
the parameters $M_5 = 2.5$ TeV, $\kappa = 1$ GeV.}
\label{fig:muons_2.5TeV_1GeV}
\end{center}
\end{figure}

In Figs.~\ref{fig:muons_1.8TeV_0.1GeV}--\ref{fig:muons_2.5TeV_1GeV}
our predictions for the process $e^+e^- \rightarrow \mu^+ \mu^-$ are
shown.

One should conclude that at fixed value of the gravity scale $M_5$,
the cross section ratio is very sensitive even to the slight
variations of the curvature $\kappa$ (compare, for instance,
Figs.~\ref{fig:bhabha_1.8TeV_0.9GeV},
\ref{fig:bhabha_1.8TeV_0.994GeV} and \ref{fig:bhabha_1.8TeV_1GeV},
as well as Figs~\ref{fig:muons_1.8TeV_0.1GeV} and
\ref{fig:muons_1.8TeV_1GeV}). The ratio
$\sigma(\mathrm{SM+grav})/\sigma(\mathrm{SM})$ can reach tens at
some values of the parameters (see
Figs.~\ref{fig:bhabha_1.8TeV_0.994GeV},
\ref{fig:muons_1.8TeV_1GeV}).

Such a behavior of the cross section ratios comes from the explicit
form of the function $\mathcal{S}(s)$~\eqref{main_formula_final}.
Indeed, at $\eta\,(\sqrt{s}/\bar{M}_5)^3 \ll 1$ the parameter
$\varepsilon$ in Eq.~\eqref{main_formula_final} is much less than 1,
and the function $\mathcal{S}(s)$ has a significant real part.%
\footnote{Remember that $\eta(\sqrt{s}/\bar{M}_5)^3 > 1$ in the
trans-Planckian region $\sqrt{s} > 3\bar{M}_5$.}
There is an interference with the SM contributions, that results in
a nontrivial dependence of
$\sigma(\mathrm{SM+grav})/\sigma(\mathrm{SM})$ on $\cos \theta$.
Given the approximation with zero real part is used
\eqref{KK_sum_zero_widths}, no interference terms exist, and $\cos
\theta$-dependence of the cross section ratio becomes similar for
all sets of the parameters (dashed curves in
Figs.~\ref{fig:bhabha_1.8TeV_0.1GeV}--\ref{fig:muons_2.5TeV_1GeV}).

If the parameter $A = \sqrt{s}/\kappa + \pi/4$ in formula
\eqref{main_formula_final} obeys the equation $\cos A = 0$, the
denominator in \eqref{main_formula_final} becomes small, and,
correspondingly, the function $\mathcal{S}(s)$ becomes large. It
results in the rapid variation of the gravity contribution near
corresponding values of $\kappa$.

\begin{figure}[h]
\begin{center}
\resizebox{.6\textwidth}{!}{\includegraphics{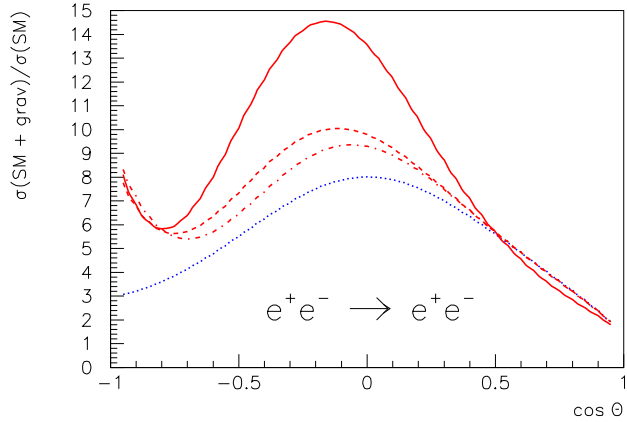}}
\caption{The dashed and dash-dotted lines correspond to smearing
of gravity cross section in the interval $(\sqrt{s} - \Delta
\sqrt{s}, \sqrt{s} + \Delta \sqrt{s})$, with $\Delta \sqrt{s} =
10$ GeV and $\Delta \sqrt{s} = 50$ GeV, respectively, while the
solid line corresponds to a case without energy smearing. The
average energy $\sqrt{s} = 1$ TeV, the parameters are the same as
in Fig.~\ref{fig:bhabha_1.8TeV_1GeV}. The dotted line is obtained
in zero width approximation The number of resonances which lie
within the energy resolution is equal to 6 (31) for $\Delta
\sqrt{s} = 10$ GeV ($50$ GeV).}
\label{fig:bhabha_average}
\end{center}
\end{figure}

\begin{figure}[h]
\begin{center}
\resizebox{.6 \textwidth}{!}{\includegraphics{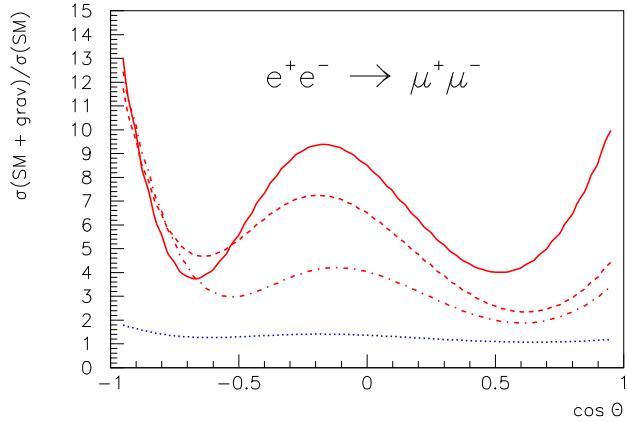}}
\caption{The same as in Fig.~\ref{fig:bhabha_average}, but for the
dimuon production with the parameters taken from
Fig.~\ref{fig:muons_1.8TeV_1GeV}. The number of resonances which
lie within the energy resolution coincides with that in the
previous figure.}
\label{fig:muons_average}
\end{center}
\end{figure}

All calculations (except for Fig.~\ref{fig:LEP_1.8TeV_0.1GeV})
were done for the fixed energy $\sqrt{s}=1$ TeV. However, the
effects related with non-zero graviton widths remain significant
after energy smearing within some interval around this value
of $\sqrt{s}$,%
\footnote{Although smaller than those without energy smearing.}
as one can see in Figs.~\ref{fig:bhabha_average},
\ref{fig:muons_average}. Note that the energy smearing has a small
influence on the cross section ratios for both processes in the
region $\cos \theta < -0.8$.

It is worth to note that both a discrete character of the mass
spectrum and nonzero widths of the KK gravitons are also important
in a case of \emph{flat} compact extra dimension, as it is shown
in Appendix.

%%%%%%%%%%%%%%%%%%%%
% Acknowledgements %
%%%%%%%%%%%%%%%%%%%%

\section*{Acknowledgements}

I am grateful to the High Energy, Cosmology and Astroparticle
Physics Section of the ICTP, where the present work was completed,
for hospitality. I thanks G.F.~Giudice and V.A.~Petrov for fruitful
discussions.

%%%%%%%%%%%%
% Appendix %
%%%%%%%%%%%%

\setcounter{equation}{0}
\renewcommand{\theequation}{A.\arabic{equation}}

\section*{Appendix A}
\label{app:A}

Let us demonstrate that the account of the graviton widths changes
zero width result by considering a simpler case of one extra flat
dimension~\cite{Arkani-Hamed:98}. The contribution of virtual
$s$-channel KK gravitons is then defined by the quantity
\begin{equation}\label{ADD_KK_sum}
\tilde{\mathcal{S}}(s) =  \frac{1}{\bar{M}_{\mathrm{Pl}}^2}
\sum_{n=1}^{\infty} \frac{1}{s - m_n^2 + i \, m_n \Gamma_n} \; .
\end{equation}
The reduced Planck mass is related with a 5-dimensional reduced
gravity scale, $\bar{M}_{4+1}$, by the relation:
\begin{equation}\label{ADD_hierarchy_relation}
\bar{M}_{\mathrm{Pl}}^2 = 2 \pi \bar{M}_{4+1}^3 R_c \; ,
\end{equation}
with $R_c$ being the radius of the extra dimension. Remember that
$\bar{M}_{4+1} = (2 \pi)^{-1/3} M_{4+1}$~\cite{Giudice:99}.

The masses of the KK gravitons are
\begin{equation}\label{ADD_graviton_masses}
m_n = \frac{n}{R_c}, \qquad n=1,2 \ldots\; ,
\end{equation}
while the graviton widths are given by~\cite{Han:99}%
\footnote{For large masses $m_n \sim \sqrt{s}$ which make the
leading contribution to the sum~\eqref{ADD_KK_sum}.}
\begin{equation}\label{ADD_graviton_widths}
\Gamma_n = \eta \, \frac{m_n^3}{\bar{M}_{\mathrm{Pl}}^2} \; .
\end{equation}

The function $\tilde{\mathcal{S}}(s)$~\eqref{ADD_KK_sum} can be
rewritten as
\begin{equation}\label{ADD_sum}
\tilde{\mathcal{S}}(s) = \frac{R_c^4}{i \eta} \, \frac{1}{b^2 -
a^2} \sum_{n=1}^{\infty} \left( \frac{1}{a^2 - n^2} - \frac{1}{b^2
- n^2} \right)\; ,
\end{equation}
where
\begin{align}
a &= \sqrt{s} R_c + \frac{i \eta}{2} \, \frac{s^{3/2}
R_c}{\bar{M}_{\mathrm{Pl}}^2} \label{a}  \; , \\
b^2 &= -i \, \frac{\bar{M}_{\mathrm{Pl}}^2 R_c^2}{\eta} \; .
\end{align}
Note that $|a| \simeq \sqrt{s} \, R_c \gg 1$, and $|b| \gg |a|$.
In the region $\sqrt{s} \sim \bar{M}_{4+1} \ll
\bar{M}_{\mathrm{Pl}}$, the second term in the sum \eqref{ADD_sum}
is non-leading, and we get (with negligible corrections of the
type $\mathrm{O}(\sqrt{s}/\bar{M}_{\mathrm{Pl}})$ omitted):
\begin{equation}\label{ADD_sum_asym}
\tilde{\mathcal{S}}(s)\simeq \frac{R_c^2}{\bar{M}_{\mathrm{Pl}}^2}
\, \sum_{n=1}^{\infty} \frac{1}{a^2 - n^2}\; .
\end{equation}

The sum in Eq.~\eqref{ADD_sum_asym} is known to be
\begin{equation}\label{n_sum}
\sum_{n=1}^{\infty} \frac{1}{a^2 - n^2} = \frac{\pi}{2 a} \,
\cot(\pi a) - \frac{1}{2 a^2} \; .
\end{equation}
By using formulae
\begin{align}\label{sin_cos}
\cos  (x+iy) &= \cos x \cosh y - i  \sin x \sinh y \; ,
\nonumber \\
\sin (x+iy) &= \sin x \cosh y + i \cos x \sinh y \; ,
\end{align}
we get from Eqs.~\eqref{ADD_sum_asym}-\eqref{sin_cos} and
hierarchy relation \eqref{ADD_hierarchy_relation}:
\begin{equation}\label{ADD_formula}
\tilde{\mathcal{S}}(s) = \frac{1}{8 \bar{M}_{4+1}^3 \sqrt{s}} \;
\frac{\sin 2B - i \sinh 2 \delta }{\sin^2 \! B + \sinh^2 \! \delta
} \; ,
\end{equation}
where
\begin{equation}\label{parameter_B_delta}
B = \pi R_c  \sqrt{s} \; , \qquad \delta  = \frac{\eta}{4} \Big(
\frac{\sqrt{s}}{\bar{M}_{4+1}} \Big)^3 \; .
\end{equation}

As one can see from \eqref{ADD_formula}, a magnitude of the
virtual graviton contribution is defined by the TeV scale
$\bar{M}_{4+1}$, not by the Planck mass $\bar{M}_{\mathrm{Pl}}$.
Moreover, the formula~\eqref{ADD_formula} is similar to formula
\eqref{main_formula_final} up to replacements $\bar{M}_5
\rightarrow 2^{1/3} \bar{M}_{4+1}$, $\kappa \rightarrow 1/(\pi
R_c)$.

The real part of $\tilde{\mathcal{S}}(s)$ in
Eq.~\eqref{ADD_formula} gives a small contribution after energy
smearing due to its rapid oscillations in $\sqrt{s}$. The
imaginary part of $\tilde{\mathcal{S}}(s)$ in
Eq.~\eqref{ADD_formula} has a correct zero width limit ($\eta
\rightarrow 0$), as one can easily check with the use of
well-known formula
\begin{equation}\label{delta_function}
\frac{1}{\pi} \, \lim_{\epsilon \rightarrow 0} \frac{\epsilon}{x^2
+ \epsilon^2} = \delta(x) \; .
\end{equation}
It also coincides with zero width expression in the
trans-Planckian energy region, namely, at $\sqrt{s} >
\bar{M}_{4+1} (4/\eta)^{1/3} \simeq 3.5 \bar{M}_{4+1}$.

However, at $\sqrt{s} \lesssim 3.5 \bar{M}_{4+1}$ the exact
formula~\eqref{ADD_formula} and zero width formula result in
different predictions by analogy with the RS scenario which has
been considered in the present paper.

Let us underline that the so-called zero width formula is obtained
under assumption that a set of the graviton resonances can be
replaced by a continuous mass spectrum:
\begin{align}\label{ADD_graviton_propagator}
\tilde{\mathcal{S}}(s) &=  \frac{1}{\bar{M}_{\mathrm{Pl}}^2}
\sum_{n=1}^{\infty}
\frac{1}{s - m_n^2 + i \, m_n \Gamma_n} \nonumber \\
&\rightarrow \frac{1}{\bar{M}_{\mathrm{Pl}}^2} \sum_{n=1}^{\infty}
\frac{1}{s - m_n^2 + i0} \nonumber \\
&\rightarrow -\frac{i \pi R_c}{\bar{M}_{\mathrm{Pl}}^2}
\int\limits_0^{\infty} \! dm \, \delta (s - m^2) =  -\frac{i}{4
\bar{M}_{4+1}^3 \sqrt{s}} \; .
\end{align}
The last step in \eqref{ADD_graviton_propagator} is justified if
only
\begin{equation}\label{inequality}
\frac{1}{R_c} < \Gamma_n
\end{equation}
is satisfied for relevant KK modes ($n \sim \sqrt{s}\, R_c$, in
our case). This inequality is equivalent to the inequality
$\sqrt{s} > \bar{M}_{4+1} (2 \pi/\eta)^{1/3} \simeq 4
\bar{M}_{4+1}$ already derived above.

Thus, one has to conclude that the approximate expression for
$\tilde{\mathcal{S}}(s)$ \eqref{ADD_graviton_propagator} can be
used only at $\sqrt{s} > 4
\bar{M}_{4+1}$.%
\footnote{When neighboring KK resonances overlap, since their
widths become lager than the mass splitting.}
In the kinematical region $\sqrt{s} \lesssim 4 \bar{M}_{4+1}$, the
gravitons widths should be taken into account, and the KK
sum~\eqref{ADD_KK_sum} \emph{cannot} be replaced by integration in
graviton masses.

%%%%%%%%%%%%%%
% References %
%%%%%%%%%%%%%%

\vfill
\eject

%%%%%%%%%%%
% Figures %
%%%%%%%%%%%

%%%%%%%%%%%%%%%%%%%%%
% Bhabha scattering %
%%%%%%%%%%%%%%%%%%%%%

%%%%%%%%%%%%%%%%%%%%%%%%%%%%%%%%%%%%
% e+e- annihilation into two muons %
%%%%%%%%%%%%%%%%%%%%%%%%%%%%%%%%%%%%

%%%%%%%%%%%%%%%%%%%%%%%%%%%%%%%%%%%%%%%%%%%%
% Energy smearing of gravity cross section %
%%%%%%%%%%%%%%%%%%%%%%%%%%%%%%%%%%%%%%%%%%%%

\end{document}